\documentclass[runningheads]{llncs}
\usepackage{graphicx}

\begin{document}

\pagestyle{headings}

\mainmatter
\title{Genetic Paralog Analysis and Simulations}

\titlerunning{Genetic Paralog Analysis and Simulations}

\author{Stanis{\l}aw Cebrat\inst{1}
\and Jan P. Radomski\inst{2}
\and Dietrich Stauffer\inst{3}
}

\authorrunning{Cebrat, Radomski, Stauffer}

\institute{Institute of Genetics and Microbiology,
University of Wroclaw,\\ ul. Przybyszewskiego 63/77, PL-54148 Wroclaw, Poland \\
\email{cebrat@microb.uni.wroc.pl}\\
\and
Interdisciplinary Center for Computational and Mathematical Modeling, \\
Warsaw University, PL-02-106 Warsaw, Poland
\and
Institute for Theoretical Physics, Cologne University, D-50923 K\"oln, Euroland
}

\maketitle

\begin{abstract}
Using Monte Carlo methods, we simulated the effects of bias in generation
and elimination of paralogs on the size distribution of paralog groups. It
was found that the function describing the decay of the number of paralog
groups with their size depends on the ratio between the probability of
duplications of genes and their deletions, which corresponds to different
selection pressures on the genome size. Slightly different slopes of curves
describing the decay of the number of paralog groups with their size were
also observed when the threshold of homology between paralogous sequences
was changed. 
\end{abstract}

\section{Introduction}
It is widely accepted that evolution is driven by two random processes -
mutations and recombinations and a directional process - selection.
Recombination not only re-shuffles
genes inside genomes or between genomes but it is also responsible for
amplification or elimination of sequences. Duplication of complete coding
sequences produces additional copies of genes called paralogs. Thus,
paralogous genes are homologous sequences arisen through gene duplication
and parallel evolution in one genome. Paralogs can also appear by
duplication of large fragments of chromosomes or even by fusion of
different genomes (allopolyploidization). Before the fusion, corresponding
sequences in the two genomes which had a common ancestor in the past are
called orthologs [1]. Since it would be very difficult to reproduce their real
history, when they appear in the genome of one organism they are
recognized as paralogs.
Paralogs are a source of simple redundancy of information, making the
genome more stable and resistant to mutational effect by complementing the
function of one copy when the other is  damaged by mutation [2] or by
reinforcing the function of the amplified gene. Most importantly,
gene duplication generates a sequence with a defined function but released
from the selection pressure. Redefinition of the duplicated gene function
may ameliorate the biological potential of the individual. Taking under
consideration all the profits brought by paralogs one can ask why the
number of paralogs seems to be limited. First of all, a higher number of
gene copies, frequently causing a higher level of products does not mean a
more concerted expression of the gene function. The best example - the
Down syndrome - is caused by redundant information. Second, limitation
comes also from the cost of replication and translation of information,
which leads to selection pressure on the genome size. The genome size
is the result of compromise between the trends for accumulating
information and keeping the costs of replication in the reasonable limit.
Nevertheless, the level of redundancy in genetic information is high, for
example in a uni-cellular eukaryote organism - Saccharomyces cerevisiae
(baker's yeast) - probably no more than 20 \% of genes fulfill essential
functions and stay in unique copies. The function of the rest of genes can
be complemented, probably mostly by paralogous sequences [3], [4].

According to the definition,
all the genes in the genome which have a common ancestor belong to one
paralog family or group. However, the genome analysis does not give us
direct information about the descent of sequences from the common ancestor
because we can only conclude about the common progenitor on the basis of
homology between compared currently "living" sequences. The level of
homology could additionally indicate the time when the two sequences have
diverged. Approximately, the number of mutations which have occurred in
the diverging sequences grows with time linearly, though it may depend on
the topological character of the duplication itself (i.e. duplication with
or without inversion)[7]. Furthermore, the fraction of positions in which the
two sequences differ does not grow linearly because of multiple
substitutions (substitutions which have occurred in the same position
several times) and reversions whose probability grows in time.
Thus, the level of homology is not an exact measure of
divergence time (branching time). At large time distances the homology
between two paralogs could be too low to recognise properly whether the
observed homology is accidental or the compared  sequences actually
descend from one progenitor sequence. That is why a threshold of homology
is assumed - if the homology level is below the threshold, the compared
sequences are considered as independently evolved. Since the threshold is
arbitrary, and differs in different analyses, it is important
to find whether the size distribution of paralog families depends on the
cutoff  level of homology.

In all analyzed
genomes the distribution of paralog families follows a specific rule.
Some authors claim an exponential function [5], others a power law
ruling the frequency of the occurrence of the folds or protein families
[6], [8], [9]. The latter authors assumed a
limited number of the
initial sequences evolving into the full genome of the contemporary organism.
In our simulations we have assumed that the evolution of the contemporary
genomes has started with all the genes indispensable for survival of the
individuals and these initial genes were independent progenitors of all
paralog families. The organisation of these genes in higher hierarchy
(families or folds) was neglected. We have analysed how the size
distribution of paralog families depends on the selection pressure, on
genome size and on the arbitrarily accepted threshold of homology deciding
about the grouping of the sequences into paralog families. The selection
pressure is an objective force influencing the genome evolution while the
paralog identification errors are connected with our ignorance, rather.
In our simulations we used two different ways for measuring the distance
between paralogs: the first one was somewhat absolute because it measured
the real time of duplication and the second one corresponded to the
homology analysis - the Hamming distance between two sequences (bit-strings)
was measured.

\section{Experimental Distributions}
Analysis of the first completely sequenced genome demonstrated  that
distributions of sizes for paralog families indicate a high level of gene
duplication [10]. Initial comparison, of bacterial, archeal
and eukaryotic genomes has shown that the number of sequences in protein
families vs. corresponding family sizes displays power law distributions [8,11].

In contrast, Slonimski et al. [5] in an one page note, reported
that for protein families of $N = 2$ to $5-6$ members, the clusters of
$N+1$ contain half the number of proteins observed in clusters of $N$,
independently of the microbial genome size. Their methodology [12], [13]
used Smith-Waterman scores $SW > 22$, the Z-significance
values, and connective-clusters in which a given sequence had similarity
of $Z_{value} \ge 8$ with one or more other sequences. The analysis have been
performed on yeast and 4 microbial genomes.

Yanai et al. [9] have compared paralog distributions for 20 genomes, using
BLAST and E-significance values ranging from
$E = 10^{-10}$ to as large as $E = 10^{-3}$.
They report linear fits on log-log scale for all genomes, with somewhat
noisy behaviour for larger families.
Qian et al. [14] have linked the power law distribution of gene families in
genomes, with the distribution of structural motifs and protein folds, all
three displaying identical slope on log-log plots. Their analyses involved
again 20 microbial genomes, and also inter-genome comparisons within analogous
functional and structural families.

Unger et al. [15] compared orthologous gene
distributions in three large curated databases: COG, ProtoMap, and Predom
(28031, 81286 and 278584 sequences respectively), and also performed partial
analysis of a human genome. They again observed a power law behaviour relating
the number of sequences in structural and functional families $F(N)$ of a given
size $N$, by $F(N) \propto N^{-b}$,  where $b$ - the slope of linear fits on 
log-log plots. Additionally they have linked the slopes for
small families, and those for large families by $b_{50} = 1 + 1/b_{500}$, where
$b_{50}$ and $b_{500}$ stand for the 50 smallest, and
the 500 largest families, after ranking them by size.

Nimwegen [16] has observed power laws, comparing the number of genes in
functional categories vs. total number of genes in a genome, 
with exponents varying both between bacterial, archeal and eukaryotic
genomes, and especially between functional categories: from 0.13 for the
protein synthesis in bacteria, to as high as 3.36 for the defense response in
eukaryotes.

\begin{figure}[hbt]
\begin{center}
\includegraphics[angle=-90,scale=0.35]{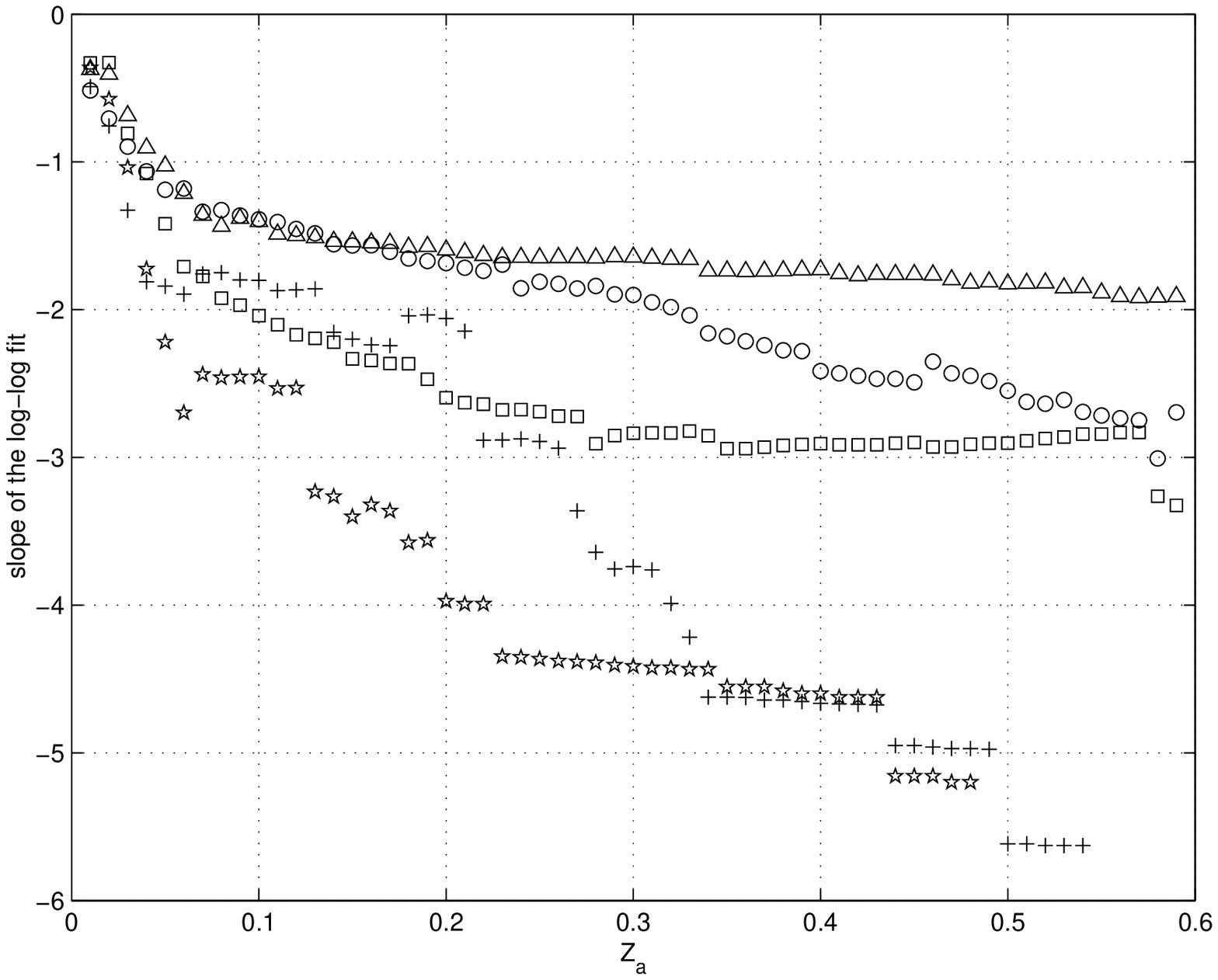}
\includegraphics[angle=-90,scale=0.35]{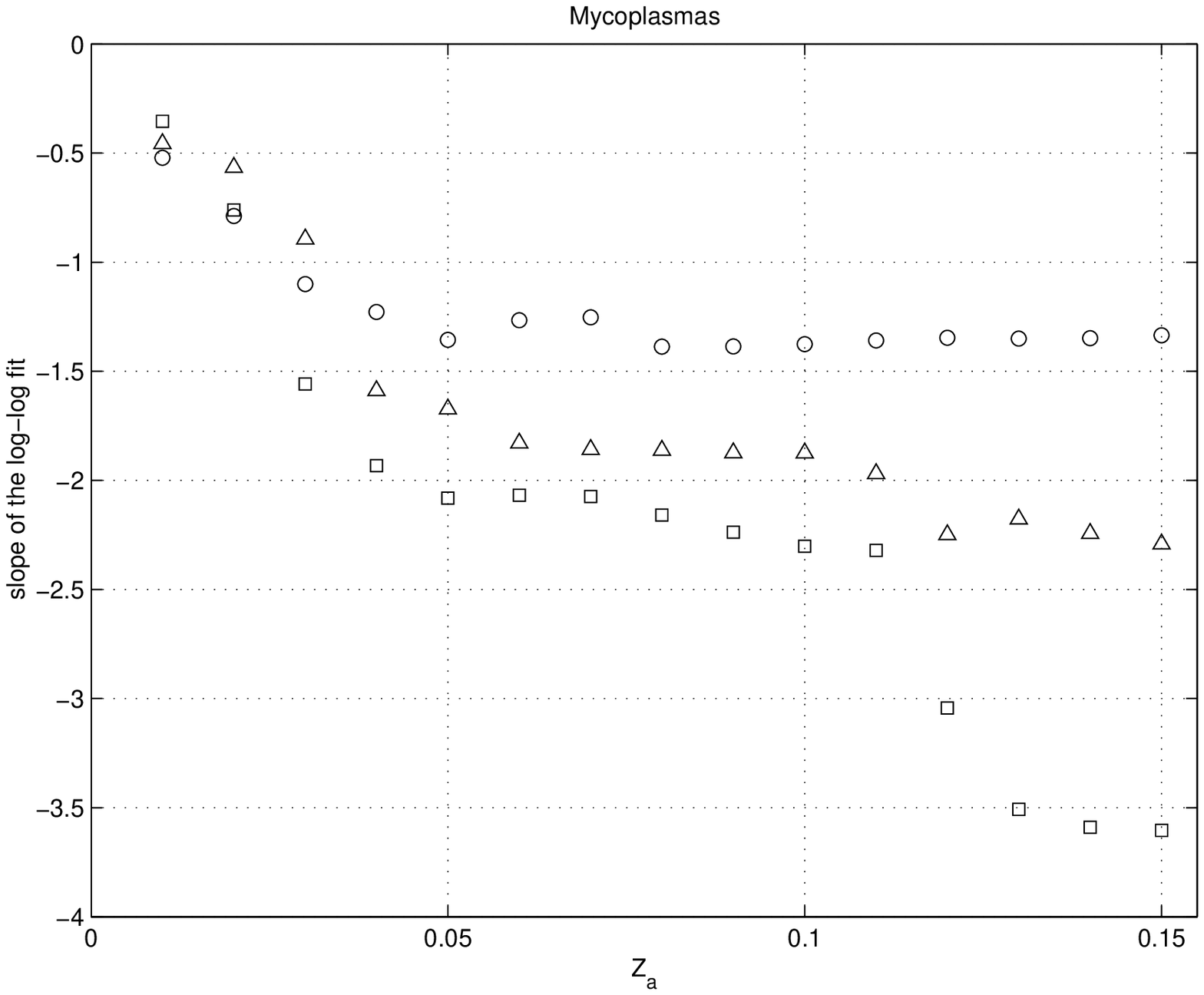}
\end{center}
\caption{
Part a: Slopes of the log-log fittings as a function of the
$Z_a$ cut-off values. {\it Borrelia burgdorferi} - pentagrams (850 genes);
{\it Haemophilus influenzae} - crosses (1712 genes); {\it Metanococcus
jannaschii} - squares (1721 genes), {\it Sulfolobus solfataricus} - triangles
(2939 genes), {\it Arabidopsis thaliana} - circles (26462 genes). \quad 
Part b: Dependence of the power law exponents on genome size for three {\it Mycoplasma}
bacteria: {\it M.genitalium} (486 genes) - squares, {\it M.pneumoniae} (687
genes) - triangles, and {\it M.pulmonis} (778 genes) - circles
}
\end{figure}

\begin{figure}[hbt]
\begin{center}
\includegraphics[angle=-90,scale=0.45]{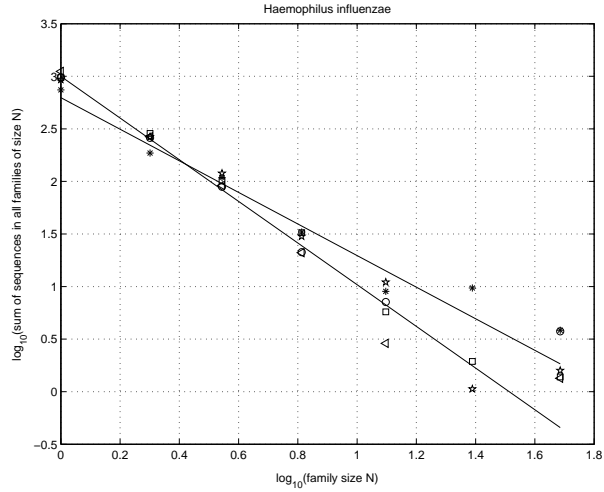}
\end{center}
\caption{
 Comparison of the log-log plots for {\it Haemophilus influenzae}
between the data from Brenner et al. [10] - circles; the current work:
$Z_a = 0.04$ - stars, $Z_a = 0.5$ - squares, and $Z_a = 0.06$ - triangles;
and when using [12],[13] $Z_{value} \ge 8$ - pentagrams. The steeper solid
line, for fitting all but the last point of Brenner's data, has the slope --2,
the more shallow one the slope of --1.5 uses all points. All three methods
are based on the use of significance for the Smith-Waterman local alignments.
}
\end{figure}

\section{Current Work}

The $Z_{value}$ [12, 13] data of all intragenomic pairwise alignments for 61
complete genomes [18] have been used. In no case an exponential decay
for a distribution of paralogous family sizes was found, independently of
the cut-off threshold of the $Z_{value}$ used as a similarity measure.
As the $Z_{value}$ depends much on the length of compared sequences [12, 13],
here we use an amended similarity measure between sequences $A$ and $B$,
$Z_a = Z_{value}(A,B) /\max[Z_{value}(A),Z_{value}(B)]$.
For identical sequences
$Z_a = 1$, and it tends to zero with increasing dissimilarity.

Figure 1a presents the slopes of the log-log fittings as a function of the
$Z_a$ cut-off values between 0.01 and 0.6 used, for several genomes.
For $Z_a < 0.04 - 0.05$,
for all genomes there are only one or two huge super-clusters, and small
fractions of singletons and doublets (sometimes also triplets). Clearly
such a small cut-off is too low to distinguish anything of interest.
For high values of $Z_a$, but obviously depending
much on the genome size, most sequences are similar only to themselves, and
there are mostly singletons, with few still remaining doublets/triplets.
At the less stringent similarity cut-off there are regions of gradual change,
interspersed by sharp changes in behaviour - corresponding obviously
to the splitting events, when clusters are broken, and a possible relationship
between homology and function within family/cluster is disrupted. Somewhere
in between these two extremes there is a small region of usefulness, when
the slope of the log-log fits seems to depend more or less linearly on
the cut-off $Z_a$ value. Tentatively it might be attributed to a $Z_a$ range
of 0.04 - 0.1, as for most genomes analysed,  we can see a relative plateau
of the log-log slope changes with increasing $Z_a$.

Moreover, as can be seen in Fig. 1a, any comparisons between genomes
must depend to a high degree on the cut-off value of the similarity
measure actually used. For example, the data of Brenner et al. [10]
for $Haemophilus$ $influenzae$ would suggest the slope of the log-log
plot equal --1.50, which would imply, if compared to the Fig. 1a, the $Z_a$ in
between of 0.02 and 0.04, clearly in a twilight zone before the supposedly
useful region of linear dependence of the slope on $Z_a$. However,
the last point (Fig. 2, circles) changes
the slope of the fit significantly, the slope after its exclusion equals
--1.98. The corresponding analysis using $Z_a$ reveals (Fig. 2, stars
$Z_a = 0.04$, squares $Z_a = 0.05$, triangles $Z_a = 0.06$) that best
agreement between ref. 10 and
the current work is at $Z_a = 0.05$, and that in both cases power law
approximation underestimates big-sized families (rightmost points, Fig. 2),
especially at higher $Z_a$. Finally, the results of
cut-off $Z_{value} = 8$, used by Slonimski et al. [5], [12] (Fig. 2,
pentagrams), again agree with both Brenner's and current results.

The often emphassized dependence of the fitted log-log slopes on the genome 
size can be observed only as a general
trend, with many exceptions. {\it Metanococcus janaschii}
and {\it Haemophilus influenzae} are of almost identical size of about 1700
sequences, but their behaviour is strikingly different, with {\it H. influenzae}
displaying the quickest change of slopes with increasing $Z_a$ of all genomes
analysed. Also, {\it H. influenzae} large clusters are breaking down to
singletons much faster (e.g. the rightmost crosses of {\it Haemophilus} in
Fig. 2, correspond to the bipartite composition
of vast majority of singletons, and a very small remainder of what
was before one or two big families).
{\it Sulfolobus solfataricus} - at approximately
one tenth the genome size of {\it Arabidopsis thaliana} - shows the most
shallow dependence of slopes on $Z_a$ of all genomes under study, comparable to
that of {\it Arabidopsis}. Even for the smallest genomes (Fig. 1b) of {\it Mycoplasma
genitalium} (486 genes), {\it Mycoplasma pneumoniae} (687 genes), and
{\it Mycoplasma pulmonis} (778 genes), Fig. 1b (squares, triangles, and circles
respectively), which because of their taxonomical proximity can be compared
directly relatively easy, the size dependence of the power law exponent
is rather perturbed.

\begin{figure}[hbt]
\begin{center}
\includegraphics[angle=-90,scale=0.35]{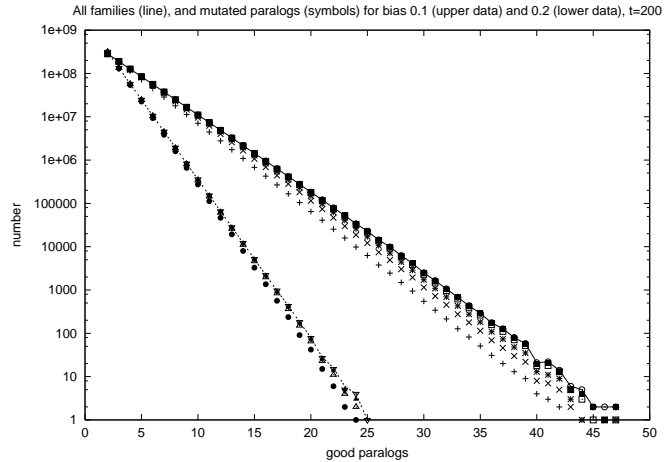}
\end{center}
\caption{
Line shows the number $n_k$ of families with $k$ paralogs each,
independent of the bit-string status. The symbols give, for $x$ = 0,1,2,4,,,,
from bottom to top, the normalized number of paralog pairs within such families
of size $k$.  $p_{mut} = 0.01, \; b = 0.1$ (upper data) and 0.2 (lower data).
}
\end{figure}

\begin{figure}[hbt]
\begin{center}
\includegraphics[angle=-90,scale=0.35]{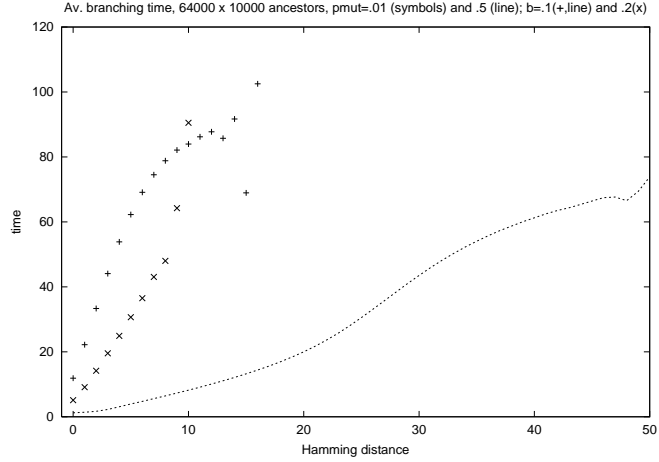}
\end{center}
\caption{
Average branching time, defined as the number of iterations since the last
creation of the paralog, versus Hamming distance, from 64,000 samples of 10,000
ancestors each, with 200 iterations. The fluctuations in this time are about as
large as the average.  Plus signs: $p_{mut} = 0.01, \; b = 0.1$; x crosses:
$p_{mut} = 0.01, \; b = 0.2$; line: $p_{mut} = 0.5, \; b = 0.1$.
}
\end{figure}

\section{Simulations}
The results of earlier modeling efforts can be found in Refs. 2,9,11,14-17,19.
In our simulations we return to the problem emphasized in the Introduction, the
number of paralogs for one given function or gene. Thus, in contrast to
what was described in preceding sections, we assume to know
for every part of the genome its function. In a simulation that is easy, since
we can follow the whole evolution since the beginning; for real genomes, such
knowledge in general still lies in the future. Our model is a simplification
of our earlier one [2], which was shown [19] to give reasonable ageing results.

The simulations start with $N$ bit-strings of length $L$ each, which are zero
everywhere. Then at each iteration each bit-string with mutation probability
$p_{mut}$ selects randomly one of its bits and flips it. Before that, also
at every time step, for each family (offspring of one ancestor) either the last
bit-string is deleted (with probability $1/2 + b$) or a randomly selected
bit-string is duplicated (with probability $1/2 - b$) and then becomes the last;
the positive bias $b$ keeps
the number $k$ of copies (``paralogs'') for each of the $N$ original bit-strings
limited. Also, the number $k$ is not allowed to become negative. Thus at any
time we have for each of the $N$ ancestors a family consisting of the first
bit-string and possibly $k-1$ additional copies or paralogs, amounting to $k$
bit-strings in total for one ancestor (=gene = function). 

The Hamming distance (= number of bits different in a bit-by-bit comparison
of two bit-strings) was calculated for each paralog with all other bit-strings
in the same family at the same time, giving $(k-1)k/2$ Hamming distances.

The simulations mostly used $L = 64, \; N = 10000, \; b = 0.1, \; p_{mut}
= 0.01$ for $t=200$ iterations and averaged over 64000 samples.
Simulations for $L = 8, \; 16, \; 32$ barely differed in the results when
a comparison was possible.  The average number $k$ of paralogs was nearly 3,
i.e. we had nearly two additional bit-strings (plus the first one) for each
ancestor. Semilogarithmic plots, Fig. 3,
of the number of paralog pairs for one ancestor with Hamming distance not
exceeding $x$ different bits typically gave straight lines with  slopes only
slightly depending on $x$. $x$ was taken as 0,1,2,4,8,16,32, and 64. For large
$x$ the curves nearly overlap. For
clarity we divided for our figure the number of pairs by the normalizing factor
$[k(k-1)/2]$ and thus for $L=64$ get the total number of families. 

The overall distributions $n_k$, lines in Fig. 3,
decay exponentially, proportional to $[(0.5-b)/(0.5+b)]^k = 1/1.5^k$, in the 
stationary state achieved after dozens of
iterations for $n_k$, even when the Hamming distances still grow. This formula
follows from a detailed balance condition that as many families move on average
from size $k$ to size $k+1$ as move in the opposite direction from size $k+1$
to size $k$. Thus the fraction of families with only a single bit-string is
$1 - (0.5-b)/(0.5+b) = 4b/(1+2b)$ in this geometric series.

We define the creation of a new paralog as a branching event and store this
time. At the end of the simulation we determine for each pair within each family
the last event they branched away from each other; the time between this last
event and the last iteration of the simulation is the branching time.
Within each family the branching times fluctuate strongly but their average
value for one given Hamming distance increases roughly linearly with that
Hamming distance, until for large Hamming distances the statistics becomes
poor, Fig. 4. For longer times (500 and 1000 iterations) the linearity improves.

\bigskip
The above model follows ref.2 except that no selection of the
fittest and similar complications are included now. Each of the ancestors
is interpreted as one function (or gene) in the whole organism. The bit-string
for this ancestor then records important mutations at different places within
this gene. The paralogs formed in the simulation from this ancestor all refer to
this one function. The first bit-string undergoes mutations just as its
paralogs and has the same properties except that it can never be removed. It
makes no sense to compare bit-strings for different functions; 00101001 means
something entirely different for the function ``brain'' than for the function
``hair''.  The $L$ bits of each bit-string correspond to $2^L$ possible
alleles for one function, not to $L$ base pairs.

The $N$ initial ancestors can also be interpreted as $N$ different samples
simulated for the same function; more generally, they could be $M$ different
genomes simulated for a genome of $N/M$ functions.

\section{Summary}
We presented here two different sets of plots: In the experimental section
we found power-law decay for the number of paralogs found by looking through
the whole genome. In the simulation section we found exponential decays
for the number of paralogs belonging to one known function. The latter
exponential decay agrees nicely with simple arguments based on detailed
balance; the slopes in these semilogarithmic plots (Fig.3) are
determined by our bias in favour of removal instead of addition of a paralog,
and the slopes barely depend on the cut-off parameter $x$ for the Hamming
distance. This agreement of theory with simulation also makes clear that our
results would be quite different if the bias would not be the same for
all functions.

\bigskip
{\bf Acknowledgements}

DS thanks the Julich supercomputer center for time on their Cray-T3E and M.
Dudek for help with LNCS formats.
JPR was
partially supported from the 115/E-343/S/ICM/853/2003 and 115/E-343/BW/ICM/1624/2003 grants.

\bigskip
{\bf \large References}

\parindent 0pt
[1] Fitch WM: Distinguishing homologous from analogous proteins, Syst. Zool.
{\bf 19} (1970) 99-113

[2] Cebrat S, Stauffer D: Monte Carlo simulation of genome viability with
paralog replacement. J.Appl.Genet. {\bf 43} (2002) 391-395

[3] MIPS 2002 Database, http://mips.gsf.de/proj/yeast/.

[4] Mackiewicz P, Kowalczuk M, Mackiewicz D, Nowicka A, Dudkiewicz M,
Laszkiewicz A, Dudek MR, Cebrat S: How many protein-coding genes are there
in the Saccharomyces cerevisiae genome? Yeast {\bf 19} (2002) 619-629

[5] Slonimski PP, Mosse MO, Golik P, Henaut A, Risler JL, Comet JP, Aude JC,
Wozniak A, Glemet E, Codani JJ: The first laws of genomics. Microb. Comp.
Genomics {\bf 3} (1998) 46.

[6] Koonin EV, Galperin MY: Sequence - Evolution - Function, Computational
approaches in Comparative Genomics (2003), Kluwer Academic Publishers

[7] Mackiewicz P, Mackiewicz D, Kowalczuk M, Dudkiewicz M, Dudek MR, Cebrat S:
High divergence rate of sequences located on different DNA strands in closely
related bacterial genomes.  J. Appl. Genet. {\bf 44} (2003) 561-584

[8]  Gerstein M, A structural census of genomes: Comparing bacterial,
eukaryotic, and archaeal genomes in terms of protein structure: J.Mol.Biol.,
{\bf 274} (1997) 562-574

[9]  Yanai I, Camacho CJ, DeLisi C: Predictions of gene family distributions in
microbial genomes: Evolution by gene duplication and modification. Phys. Rev.
Lett., {\bf 85} (2000) 2641-2644

[10]  Brenner SE, Hubbard T, Murzin A, Chotia C: Gene duplications in
{\it Haemo\-philus influenzae}. Nature, {\bf 378} (1995) 140

[11]  Huynen MA, van Nimwegen E: The frequency distribution of gene family sizes
in complete genomes. Mol.Biol.Evol., {\bf 15} (1998) 583-589

[12]  Codani JJ, Comet JP, Aude JC, Glemet E, Wozniak A, Risler JL, Henaut A,
Slonimski PP: Automatic analysis of large-scale pairwise alignments of protein
sequences. Methods Microbiol., {\bf 28} (1999) 229-244

[13]  Comet JP, Aude JC, Glemet E, Risler JL, Henaut A, Slonimski PP, Codani JJ:
Significance of Z-value statistics of Smith-Waterman scores for protein
alignments. Comput.Chem., {\bf 23} (1999) 317-331

[14]  Qian J, Luscombe NM, Gerstein M: Protein family and fold occurrence in
genomes: Power-law behaviour and evolutionary model. J.Mol.Biol. {\bf 313}
(2001) 673-681

[15]  Unger R, Uliel S, Havlin S: Scaling law in sizes of protein sequence
families: From super-families to orphan genes. Proteins {\bf 51} (2003) 569-576

[16] van Nimwegen E: Scaling laws in the functional content of genomes.
Trends Genet. {\bf 19} (2003) 479-484

[17] Koonin EV, Wolf Yi, Karev GP: The structure of the protein universe and
genome evolution, Nature, {\bf 420} (2002) 218-223

[18] TERAPROT project (CEA, Gene-It, Infobiogen), June 2002

[19] Alle P: Simulation of gene duplication in the Penna bit-string model of
biological ageing. Master's thesis, Cologne University 2003.

\end{document}